\documentclass[]{spie}  

\usepackage{amsmath,amsfonts,amssymb}
\usepackage{subcaption}
\usepackage{graphicx}
\usepackage{cite} 
\usepackage{epsfig}
\usepackage{amsmath}
\usepackage{nccmath}
\usepackage{amssymb}
\usepackage{mwe}
\usepackage{acro}
\usepackage{amssymb}
\usepackage{xcolor,colortbl}
\usepackage{tabularx}
\usepackage{relsize}
\usepackage{pifont}
\usepackage{booktabs} 
\usepackage{multirow}
\usepackage{multicol}
\usepackage{adjustbox}
\usepackage{float}
\usepackage{graphicx}
\usepackage{makecell}
\usepackage{tabu}
\usepackage[colorlinks=true, allcolors=blue]{hyperref}
\usepackage[capitalize]{cleveref}

\title{Cross-organ Deployment of EOS Detection AI without Retraining: Feasibility and Limitation}
   
\author[a]{Yifei Wu}
\author[b]{Juming Xiong}
\author[a]{Tianyuan Yao}
\author[a]{Ruining Deng}
\author[a]{Junlin Guo}
\author[b]{Jialin Yue}
\author[c]{Naweed Chowdhury}
\author[a,b]{Yuankai Huo}

\affil[a]{Department of Computer Science, Vanderbilt University, Nashville, TN, USA}

\affil[b]{Department of Electrical and Computer Engineering, Vanderbilt University, Nashville, TN, USA}

\affil[c]{Department of Otolaryngology, Vanderbilt University Medical Center, Nashville, TN, USA}

\authorinfo{Corresponding author: Yuankai Huo: E-mail: yuankai.huo@vanderbilt.edu}

\begin{document} 
\maketitle

\begin{abstract}
Chronic rhinosinusitis (CRS) is characterized by persistent inflammation in the paranasal sinuses, leading to typical symptoms of nasal congestion, facial pressure, olfactory dysfunction, and discolored nasal drainage, which can significantly impact quality-of-life. Eosinophils (Eos), a crucial component in the mucosal immune response, have been linked to disease severity in CRS. The diagnosis of eosinophilic CRS typically uses a threshold of 10-20 eos per high-power field (HPF). However, manually counting Eos in histological samples is laborious and time-intensive, making the use of AI-driven methods for automated evaluations highly desirable. Interestingly, eosinophils are predominantly located in the gastrointestinal (GI) tract, which has prompted the release of numerous deep learning models trained on GI data. This study leverages a CircleSnake model initially trained on upper-GI data to segment Eos cells in whole slide images (WSIs) of nasal tissues. It aims to determine the extent to which Eos segmentation models developed for the GI tract can be adapted to nasal applications without retraining. The experimental results show promising accuracy in some WSIs, although, unsurprisingly, the performance varies across cases. This paper details these performance outcomes, delves into the reasons for such variations, and aims to provide insights that could guide future development of deep learning models for eosinophilic CRS.

\end{abstract}

\keywords{Eosinophils, CircleSnake, Chronic Rhinosinusitis, Deep Learning}

\section{INTRODUCTION}
\label{sec:intro}  
Chronic rhinosinusitis (CRS) is a multifactorial disease influenced by both genetic and environmental factors contributions and is among the most common chronic illnesses globally.\cite{Bousquet2020,DAiuto2024} Patients with allergic rhinitis chronic sinusitis often experience nasal congestion, facial pressure, olfactory dysfunction, and discolored nasal drainage, which can significantly impact quality-of-life.\cite{Kennedy2024} Eosinophils (Eos), a type of white blood cell, play a critical role in the immune response in certain endotypes of CRS, with elevated Eos levels suggestive of more severe disease.\cite{AhlstromEmanuelsson2004,Kampe2011}

Currently, counting Eos in pathological tissue slices is a manual, time-consuming task for pathologists. However, deep learning models can automate this process, extracting features from images and achieving high accuracy and sensitivity through extensive data training. This reduces human intervention and uncovers complex features and patterns, improving diagnostic accuracy. Modern deep learning models also offer real-time image processing capabilities with hardware accelerators such as GPUs and TPUs, providing immediate diagnostic support in urgent medical settings and enhancing medical efficiency and patient outcomes. Additionally, these models are often designed to generalize well across different contexts within a specific image modality. Through techniques like transfer learning, these models can be adapted to new domains, significantly reducing the development time and cost for new models.

\begin{figure*}[ht]
\begin{center}

\includegraphics[width=\linewidth]{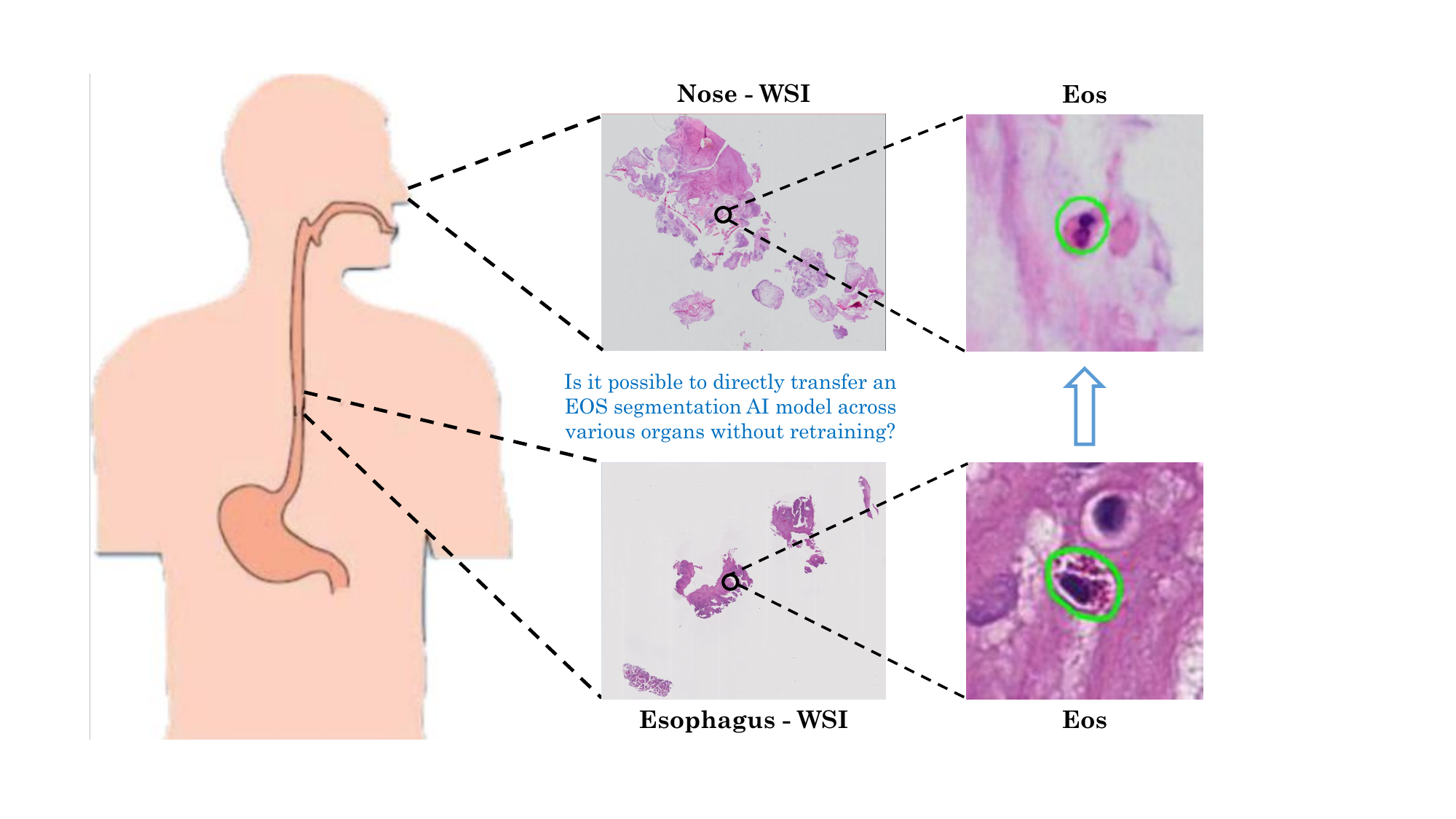}
\end{center}
\caption{The goal of this study is to determine whether and under what conditions the pre-trained CircleSnake model for eosinophil detection can be directly applied to other parts of the body, such as from the esophagus to the nose.}
\label{fig:Trans}
\end{figure*}

 Interestingly, eosinophils are predominantly located in the gastrointestinal (GI) tract, which has prompted the release of numerous deep learning models~\cite{Czyzewski2021,Daniel2021} trained on GI data. 
 For example, previous work has shown the effectiveness of the CircleSnake model~\cite{Yang2020,Nguyen2021,Xiong2024,Liu2023} in medical image segmentation tasks, particularly for eosinophilic esophagitis (EoE). CircleSnake is an end-to-end circle contour deformation-based segmentation method that has proven successful in detecting spherical medical objects.  The model, trained on segmenting Eos cells in whole slide images (WSIs) of EoE patients, achieved high accuracy and efficiency. ~\cite{Liu2023,Xiong2024}These promising results suggested the potential for applying the CircleSnake model to other medical imaging contexts, such as nasal tissues(Fig.~\ref{fig:Trans}).

In this paper, we extend the application of the CircleSnake model to the detection of Eos cells in nasal tissues, leveraging the pre-trained model developed for EoE. We aim to evaluate the model's performance in this new context and identify the specific conditions under which it excels or encounters challenges. Our analysis involves categorizing the results into three groups to illustrate the conditions under which the model performs well and when it does not, highlighting the model's limitations.

The contribution of this study is threefold:

$\bullet$ We conduct the initial study to evaluate the effectiveness of directly applying an AI developed for Eos identification in upper GI tissues to nasal tissues without any retraining.

$\bullet$ A thorough quantitative and qualitative analysis is conducted to identify specific conditions that affect the model's accuracy, such as staining intensity.

$\bullet$ This study offers insights for the future development of AI for Eos detection in nasal tissue. Given that performance fluctuates with global intensity variations, diversifying data collection and employing digital staining technology could help close the gap, leading to a more generalizable AI for Eos detection.

\section{METHOD}
\label{sec:method}  


\begin{figure*}[ht]
\begin{center}

\includegraphics[width=\linewidth]{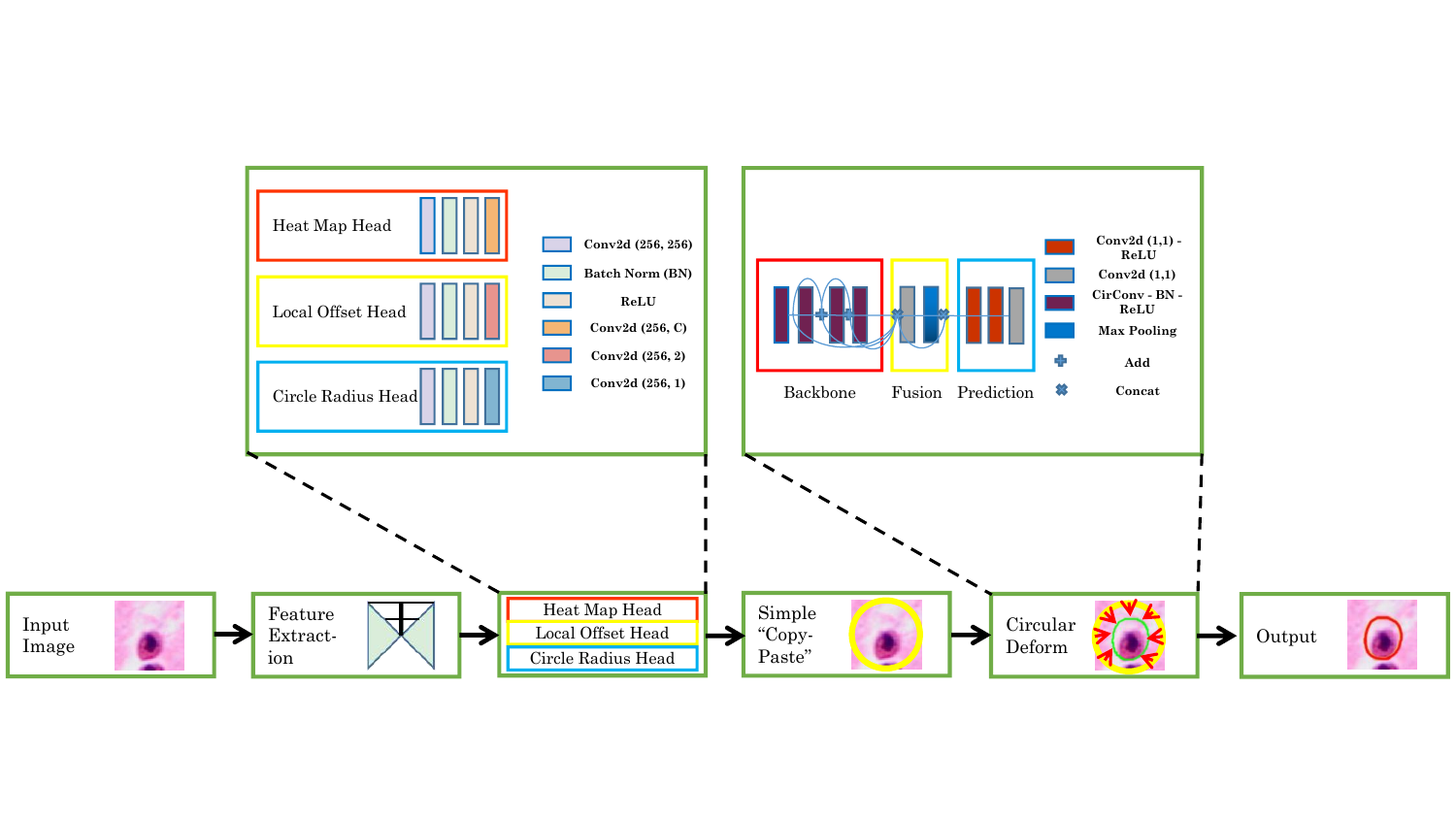}
\end{center}
\caption{A pipeline of the CircleSnake segmentation model's network structure. The initial boundary forms a circular contour after verification by CircleSnake. Through a graph convolutional network (GCN), the outline eventually transforms into the final circular boundary of the eosinophil cell.}
\label{fig:PIP}
\end{figure*}

CircleSnake~\cite{Yang2020,Nguyen2021,Xiong2024,Liu2023} has been released as a state-of-the-art method on Eos detection for the esophagus (Fig.~\ref{fig:PIP}), whose superior performance is obtained from the optimal design for detecting circle shape objects. In this study, we directly apply the esophagus-trained CircleSnake approach to segment the Eos cells on nasal tissue.  By defining an input image \(I \in \mathbb{R}^{W \times H \times 3}\) with size \(W \times H\)\cite{Zhou2019}, the Circle Proposal Localization Network (CPL Network) generates the location of the center of each object in a heat map \(\hat{Y} \in [0,1]^{\frac{W}{R} \times \frac{H}{R} \times C}\), in which \(C\) and \(R\) represents the amount of candidate class and the down-sampling factor respectively. \(\hat{Y}_{xyc}\) describes the probability the lesion center belongs to class \(c\), and \(\hat{Y}_{xyc} = 1\) stands for the center in the heat map while \(\hat{Y}_{xyc} = 0\) for the background. The target center point is presented in the form of a two-dimensional Gaussian kernel\cite{Law2018}:
\begin{equation}
Y_{xyc} = \exp \left( - \frac{(x - \tilde{p}_x)^2 + (y - \tilde{p}_y)^2}{2\sigma_p^2} \right)
\end{equation}

\noindent where \((\tilde{p}_x, \tilde{p}_y)\) is the coordinate of the down-sampled truth center points and \(\sigma_p\) represents the kernel standard deviation, and through which we can calculate the training loss \(L_k\) as\cite{Lin2017}:

\begin{equation}
       L_k = -\frac{1}{N} \sum_{xyc} \begin{cases}
(1 - \hat{Y}_{xyc})^\alpha \log(\hat{Y}_{xyc}) & \text{if } Y_{xyc} = 1 \\
(1 - Y_{xyc})^\beta (\hat{Y}_{xyc})^\alpha \log(1 - \hat{Y}_{xyc}) & \text{otherwise}
\end{cases}
\end{equation}

\noindent where \(\alpha\) and \(\beta\) are both hyper-parameters.

By selecting \(n\) top peaks whose values are no less than its neighbors from all directions, the center points are defined as \(\hat{P} = \{(\hat{x}_i, \hat{y}_i)\}_{i=1}^{\hat{n}}\). The offset \((\delta \hat{x}_i, \delta \hat{y}_i)\) can be computed from the offset prediction loss \(L_{\text{off}}\). The center coordinate and radius of the bounding circle are calculated as:

\begin{equation}
\hat{P} = \{(\hat{x}_i + \delta \hat{x}_i, \hat{y}_i + \delta \hat{y}_i)\}_{i=1}^{\hat{n}} \quad \hat{R} = \hat{R}_{\hat{x}_i, \hat{y}_i}
\end{equation}

\noindent where radius prediction \(\hat{R} \in \mathbb{R}^{\frac{W}{R} \times \frac{H}{R} \times 1}\) is optimized by:

\begin{equation}
L_{\text{radius}} = \frac{1}{N} \sum_{k=1}^{N} \left| \hat{R}_{\hat{p}_k} - r_k \right|
\end{equation}

\noindent where \( r_k \) represents the ground truth radius of object \( k \). And the overall objective for Circle Object Detection is:

\begin{equation}
L_{\text{det}} = L_k + \lambda_{\text{radius}} L_{\text{radius}} + \lambda_{\text{off}} L_{\text{off}}
\end{equation}

Following the definitions from Zhou et al\cite{Zhou2019}, \( \lambda_{\text{radius}} = 0.1 \) and \( \lambda_{\text{off}} = 1 \).

Based on the Circle object detection above, the bounding circle related to each object can be determined and the initial contour proposal will be drawn by the circle representation. To ensure the simplicity and consistency of the circle contour, it avoids some complicated deformation and extreme octagon proposals based on poles. \(N\) sample points \(\{x_i^{\text{circle}} | i=1,2,\ldots,N \}\) are selected isometrically starting at the top-most point \( x_1^{\text{circle}} \). Similarly, we sample \(N\) vertices along the boundary uniformly to generate the ground truth contour. \(N\) is set to 128\cite{Peng2020}.

Based on the \(N\)-vertices contour above, we first generate feature vectors for every vertex. Combining a vertex \( x_i^{\text{circle}} \) and its feature maps \( F(x_i^{\text{circle}}) \), we create the correlated input features \( f_i^{\text{circle}} \), and we treat the features as a one-dimensional discrete signal \( f: \mathbb{Z} \to \mathbb{R}^D \). Circular convolutional neural network is used on feature learning, as shown in the Fig.~\ref{fig:PIP}. The signal is defined as:

\begin{equation}
(f_N^{\text{circle}} * k)_i = \sum_{j=-r}^{r} (f_N^{\text{circle}})_{i+j} k_j
\end{equation}

\noindent where operator \( * \) is the standard convolution and \( k: [-r, r] \to \mathbb{R}^D \) stands for a learnable kernel function. The kernel size is set to 9 based on\cite{Peng2020}.

CircleSnake is implemented through the graph convolutional network (GCN), which includes backbone, fusion and prediction. The features are connected by eight "CirConv-Bn-ReLU" layers and circular convolutions layers "CirConv" and retransmissioned through the 1x1 convolutional and max pooling layer. Eventually, the prediction head covers three 1x1 convolutional layers and writes the offsets of the vertex. The loss function is given by:

\begin{equation}
L_{\text{iter}} = \frac{1}{N} \sum_{i=1}^{N} l_1(x_i^{\text{circle}} - x_i^{\text{gt}})
\end{equation}

\noindent where \( x_i^{\text{gt}} \) stands for the ground truth boundary point while \( x_i^{\text{circle}} \) represents the deformed contour point. The deformation is accomplished in 3 iterations, regressing all the \(N\) offsets based on\cite{Peng2020}.

\begin{figure*}
\begin{center}

\includegraphics[width=\linewidth]{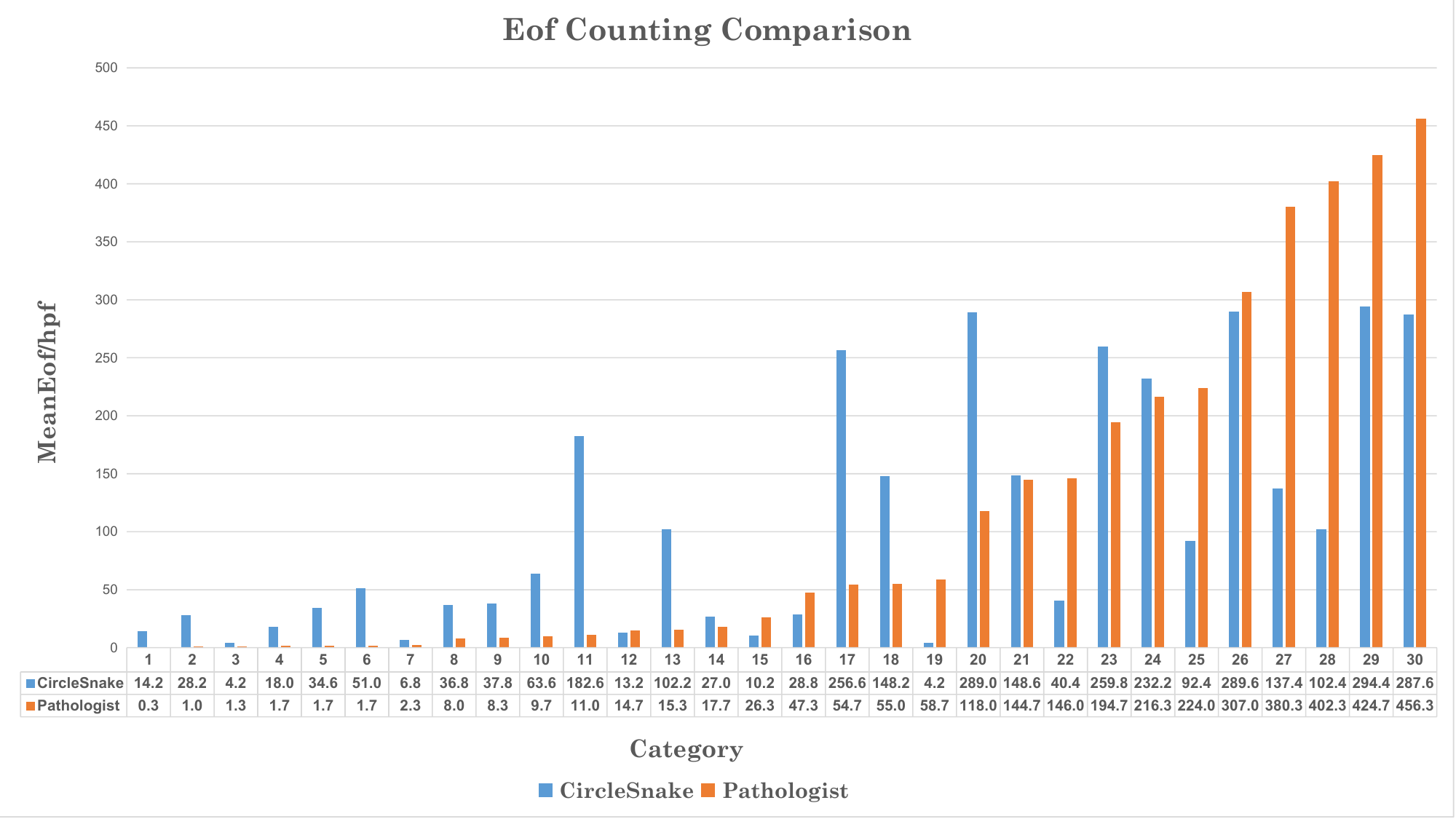}
\end{center}
\caption{The bar chart displays a comparison between counts obtained by human experts and those generated by the CircleSnake model. Each bar represents the average number of eosinophilic cells in the top five regions with the highest cell counts within a single high-power field (HPF) across all tissue sections for each patient. The CircleSnake model's data is shown in blue bars, while the data from the pathologists is depicted in orange bars. The data is arranged in ascending order based on the counts from the pathologists, while the threshold is set to 0.15.}
\label{fig:ECC}
\end{figure*}

\begin{figure*}
\begin{center}

\includegraphics[width=0.8\linewidth]{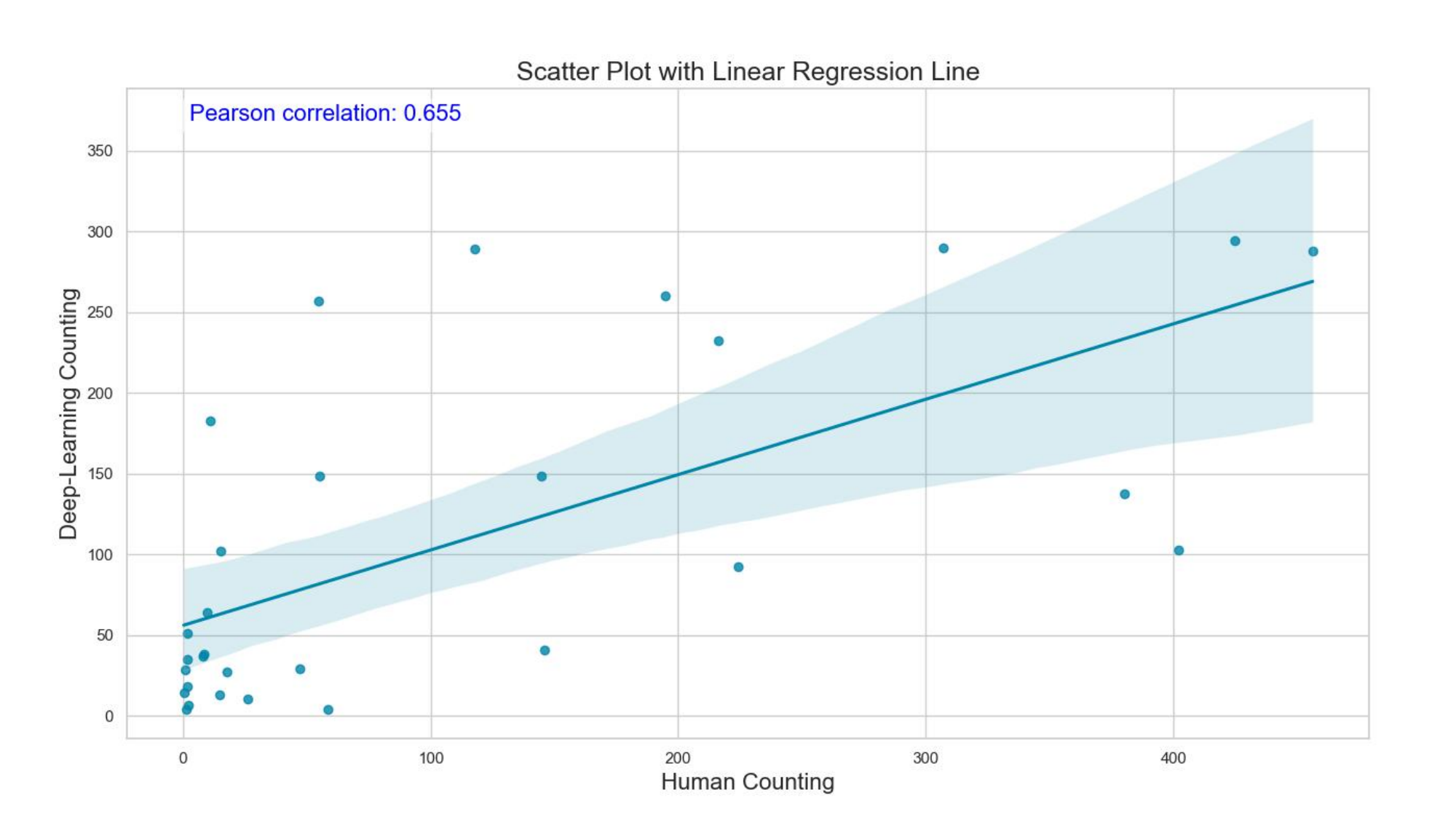}
\end{center}
\caption{
This scatter plot illustrates the comparison between human and machine counts. The blue line represents the linear regression, and the blue points indicate the coordinates of the counts, while the shaded region around the regression line depicts the confidence interval.}
\label{fig:LRL}
\end{figure*}

\section{Experimental Design}
\label{sec:experiment design}

Patients were recruited from an on-going prospective observational study enrolling CRS patients who are undergoing endoscopic sinus surgery after failing initial medical therapy for CRS consisting of topical and oral nasal steroids and high-volume saline irrigations. All patients underwent bilateral maxillary antrostomies, total sphenoethmoidectomy, and frontal sinusotomies. Patients with cystic fibrosis, ciliary dyskinesia, autoimmune diseases, or immunodeficiency were excluded from enrollment. The experimental dataset contained 200 hematoxylin and eosin-stained WSIs with 20x objective lens gaining from 30 different patients and a table recording the maximum number of Eos in one HPF in the related image and the mean value over 5 HPFs selected from the highest density regions of the WSI by a board-certified head and neck pathologist.

 Patches of 1024 x 1024 resolution are extracted from WSIs using QuPath and the meta-data information of each image was recorded into a correlated COCO-format json file. We loaded the pre-trained CircleSnake model and began the testing process which would extract the annotations on each pixel. The details of the annotations were saved into a geojson file which could be loaded in QuPath and counted by CircleSnake. 

\begin{figure*}[ht]
\begin{center}

\includegraphics[width=\linewidth]{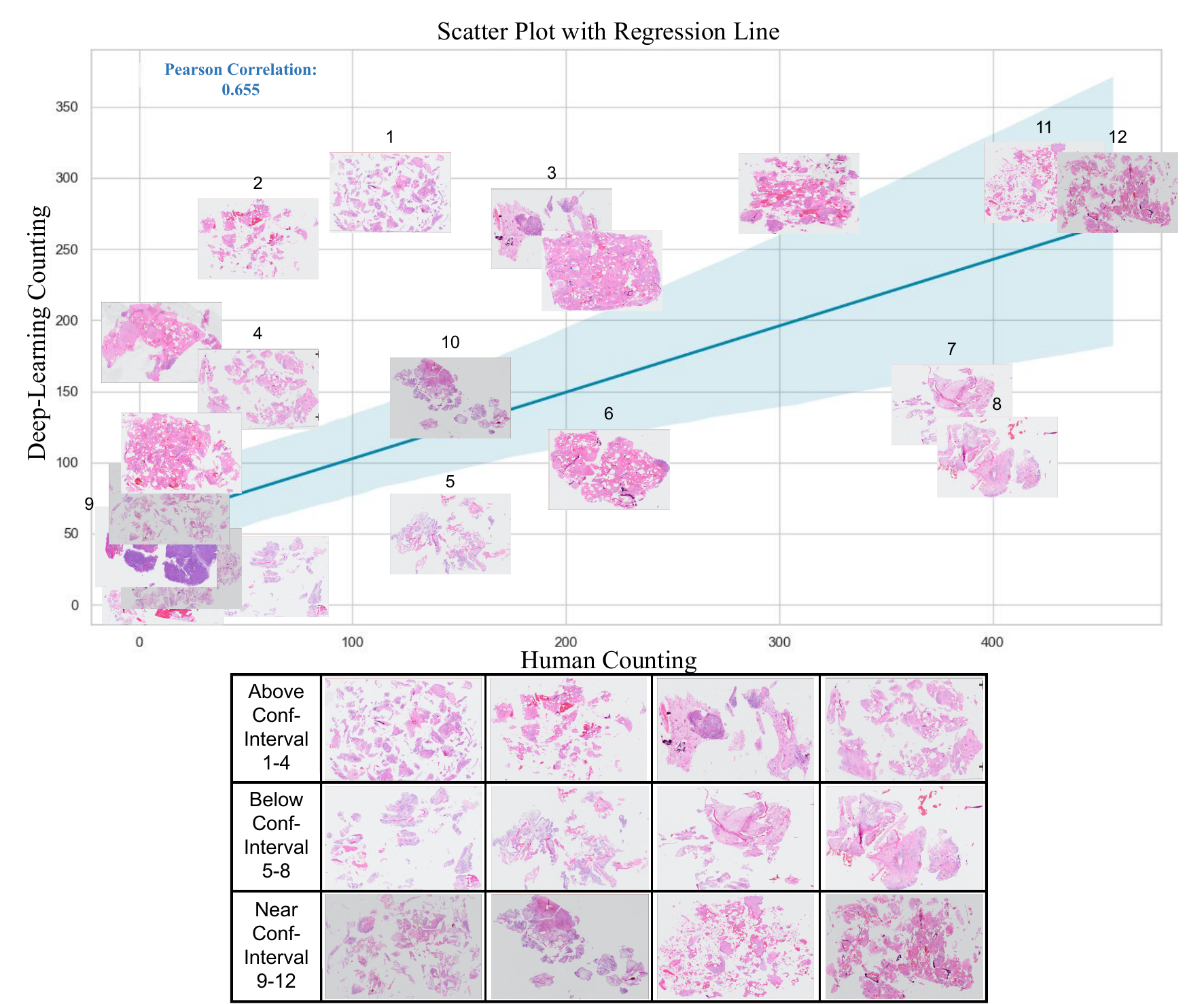}
\end{center}
\caption{This figure shows the visualization of all WSIs in the same location as Fig.~\ref{fig:LRL}. The observations and interpretation are presented in the discussion section.}
\label{fig:Group}
\end{figure*}

\section{Results}


The reading between deep learning (CircleSnake) and human experts are presented in Fig.~\ref{fig:ECC}. The quantitative correlation results are presented in Fig.~\ref{fig:LRL}, with the Pearson correlation coefficient\cite{Huo2019}. The correlation results are listed in Table.~\ref{tab:circlesnake}, showing that the r value$>$0.6 for different hyper-parmeters. Such correlations are significant (p-value$<$0.05)
For qualitative visulization, we categorized the cases into three groups: above, below, and near the confidence interval region, as shown in Fig.~\ref{fig:Group}, to discern the general pattern from each group.











\begin{table}[ht!]
\centering
\caption{Pearson Correlation Results between Human and Machine Reads.}
\begin{tabular}{>{\raggedright\arraybackslash}p{5cm}cccccccc}
\toprule
\multicolumn{9}{c}{\textbf{Pearson Correlation}} \\
\midrule
\textbf{Performance Metrics} & \multicolumn{4}{c}{\textbf{r value}} & \multicolumn{4}{c}{\textbf{p value}} \\
\cmidrule(r){2-5} \cmidrule(r){6-9}
\textbf{Threshold (ct\_score)} & 0.3 & 0.2 & 0.15 & 0.1 & 0.3 & 0.2 & 0.15 & 0.1 \\
\midrule
\textbf{Mean (of Top-5) Counting} & 0.621 & 0.653 & 0.655 & 0.627 & 0.000015 & 0.000083 & 0.000085 & 0.000108 \\
\midrule
\textbf{Maximum Counting} & 0.591 & 0.637 & 0.639 & 0.601 & 0.000580 & 0.000251 & 0.000144 & 0.000267 \\
\bottomrule
\end{tabular}
\label{tab:circlesnake}
\end{table}

\section{DISCUSSION}
This study demonstrates that the CircleSnake deep learning model, initially trained for EoE, has various performances in detecting Eos in nasal tissue scenarios. This paper performs both qualitative and quantitative analyses to assess such feasibility, and more important, the limitations.

Fig.~\ref{fig:Group} shows the visualization of all WSIs in the same location as Fig.~\ref{fig:LRL}. In this section, we summarize our observations and interpretation. First, images near the regression line (9-12) include a few medium-sized tissue regions that are relatively scattered and stained darker. Interestingly, these images seem to exhibit a pattern where the background luminance of the WSIs is substantially darker. For the first group (1-4), the tissues blocks are mostly exceedingly numerous and dispersed. For the second group (5-8), the staining is typically lighter than other groups, making a different appearance from the training data. These findings suggest that future efforts to narrow the performance gap should focus on implementing super-resolution techniques and digital stain normalization methods to minimize variations between tissue samples.

\acknowledgments 
This research was supported by NIH R01DK135597 (Huo), DoD HT9425-23-1-0003 (HCY), NIH NIDDK DK56942 (ABF) and the Burroughs Wellcome Fund (NIC). This work was also supported by Vanderbilt Seed Success Grant, Vanderbilt Discovery Grant, and VISE Seed Grant. This project was supported by The Leona M. and Harry B. Helmsley Charitable Trust grant G-1903-03793 and G-2103-05128. This research was also supported by NIH grants R01EB033385, R01DK132338, REB017230, R01MH125931, and NSF 2040462. We extend gratitude to NVIDIA for their support by means of the NVIDIA hardware grant.  
\bibliography{report} 
\bibliographystyle{spiebib} 

\end{document}